\documentclass[aps,prl,reprint,showpacs,amsmath,superscriptaddress,longbibliography]{revtex4-1}
\usepackage{amssymb}
\usepackage{mathrsfs}
\usepackage{graphicx}
\usepackage{float}
\usepackage[normalem]{ulem}
\usepackage{color}
\usepackage{bm}
\newcommand{\tabincell}[2]{\begin{tabular}{@{}#1@{}}#2\end{tabular}}

\def\blue{\textcolor{black}}

\def\GJ{\textcolor{black}}

\def\GJB{\textcolor{black}}

\begin{document}
\title{Direct dynamical characterization of higher-order topological insulators with nested band inversion surfaces}
%\red{[Maybe ``generating higher-order topological phases using nested band inversion surfaces", since there are some restriction for applying this method.]}
\author{Linhu Li}
\affiliation{Department of Physics, National University of Singapore, 117542, Singapore}
\author{Weiwei Zhu}
\affiliation{Department of Physics, National University of Singapore, 117542, Singapore}
\author{Jiangbin Gong}
\email{phygj@nus.edu.sg}
\affiliation{Department of Physics, National University of Singapore, 117542, Singapore}

\begin{abstract}
{Higher-order topological insulators (HOTIs) are systems with topologically protected in-gap boundary states localized at their $(d-n)$-dimensional boundaries, with $d$ the system dimension and $n$ the order of the topology.
This work proposes a rather universal dynamics-based characterization of one large class of $Z$-type  HOTIs without specifically relying on any symmetry considerations. The key element of our innovative approach is to connect quantum quench dynamics with nested configurations of the so-called band inversion surfaces (BISs) of momentum-space Hamiltonians as a sum of operators from the Clifford algebra (a condition that can be relaxed), thereby making it possible to dynamically detect each and every order of topology on an equal footing.  Given that experiments on synthetic topological matter can directly measure the winding of certain pseudospin texture to determine topological features of BISs,  the topological invariants defined through nested BISs are all within reach of ongoing experiments. Further, the necessity of having nested BISs in defining higher-order topology offers a unique perspective to investigate and engineer higher-order topological phase transitions.}
% thus providing a convenient scheme for experimentally detecting the topological invariants of the HOTIs.
\end{abstract}

\maketitle
%\red{Note: there are several different ``orders" in this paper. I have used $d$ for the order of spatial dimensions, $n$ %for the order of topology, and $m$ for the order of BISs. There is also the nesting order, label with $i$. Different %notations were used in previous versions, and I may have missed some during my revision.}

\section{Introduction}
Topological phases of matter feature in-gap boundary states protected by the bulk topological properties, usually characterized by  topological invariants defined throughout the Brillouin zone (BZ) \cite{hasan2010colloquium,qi2011topological}.
In the past few years, great attention has been paid to the concept of higher-order topological insulators (HOTIs)
%the notion of topological phases has been generalized by the concept of higher-order topological insulators (HOTIs)
\cite{benalcazar2017HOTI,Benalcazar2017HOTI2,benalcazar2019quantization,ezawa2018HOTI,Schindlereaat0346,song2017topological,huang2017building,fang2019new,matsugatani2018connecting,langbehn2017HOTI,song2017HOTI,ren2020engineering,erra-Garcia2018HOTI,christopher2018HOTI,imhof2018topolectrical,	schindler2018HOTI,noh2018topological,zhang2019second,xue2019acoustic,ni2019observation,fukui2018entanglement,wheeler2019many,kang2019many}, where a $n$th-order topology of a $d$-dimensional ($d$D) system manifests as robust in-gap states localized at its $(d-n)$D boundaries.
%where the system's topology manifests as nontrivial states localized at the "boundary of boundary", e.g. corners of a system in two-dimension (2D) or above.
Such ``boundary of boundary" states are fascinating because they reflect nontrivial topology
%associated to not only the $d$D bulk of the system, but also some effective lower-dimensional systems.
of both the $d$D bulk and its lower-dimensional boundaries.
The celebrated method of ``nested" Wilson loops provides a powerful topological characterization of HOTIs, which depicts higher-order topology via lower-order objects, i.e., the Wannier bands \cite{benalcazar2017HOTI,Benalcazar2017HOTI2}.
At present, topological characterization of HOTIs of different dimensions and various symmetries continues to be of great interest,
having further motivated generalization of several conventional topological invariants specifically tailored for HOTIs, such as the Zak phases \cite{liu2017novel,xie2018second}, winding numbers \cite{serra2019observation,li2018direct,seshadri2019generating,imhof2018topolectrical}, and topological indices at high-symmetric points \cite{Benalcazar2017HOTI2,ezawa2018HOTI,benalcazar2019quantization}.
Nevertheless, most of these theoretical concepts or tools have rather strong restrictions on the systems under study (e.g., orders of topology, spatial dimensions, and/or certain momentum-dependent symmetries).
%A universal topological characterization for HOTIs is still \GJnew{absent}, especially for the A and AIII classes of the Altland-Zirnbauer (AZ) symmetry classification  \cite{schnyder2008classes,ryu2010class,RevModPhys.88.035005}, the two complex symmetry classes associated with only the chiral symmetry.}
%\blue{Furthermore, direct experimental detection of topological invariants of HOTIs \GJnew{at each and every order of topology  is still lacking}, as most contemporary experimental advances have focused on detecting the spectrum and topological boundary states of HOTIs.}
Furthermore, as most contemporary experimental advances have focused on detecting the spectrum and topological boundary states of HOTIs,
a universal topological characterization of HOTIs with direct experimental detection at each and every order of topology is still lacking,
especially for the A and AIII classes of the Altland-Zirnbauer (AZ) symmetry classification \cite{schnyder2008classes,ryu2010class,RevModPhys.88.035005}, the two complex symmetry classes associated with only the chiral symmetry.

%2D Zak phase \cite{liu2017novel,xie2018second}, and boundary winding numbers \cite{serra2019observation,li2018direct}.

The starting point of this work is the recently proposed concept of band inversion surfaces (BISs) to directly probe 1st-order topology \cite{zhang2018dynamical,zhang2019dynamical,zhang2019characterizing,zhang2020unified,yu2020high}.  A BIS denotes a special region in the BZ where one or several pseudospin components of a momentum-space Hamiltonian vanishes.  By construction, the winding behavior of such pseudospin texture inside one BIS must then be different from that outside the BIS.
According to the Altland-Zirnbauer (AZ) symmetry classification \cite{schnyder2008classes,ryu2010class,RevModPhys.88.035005},
a $d$D system of the complex symmetry classes possesses a $Z$-type topology if its Hamiltonian contains $(d+1)$ anticommuting terms, forming a $(d+1)$D vector space which allows a quantized winding of the $d$D closed manifold given by the Hamiltonian vector varying throughout the BZ.
%In general, the Hamiltonian vector of a $d$D system gives a $d$D closed manifold throughout the BZ, which exhibits a $Z$-type winding topology in a $(d+1)$D vector space. Such a vector space is formed by $(d+1)$ anticommuting terms of the Hamiltonian, and the winding topology is described by a winding number or a Chern number for odd and even $d$ respectively, according to the Altland-Zirnbauer (AZ) symmetry classification \cite{schnyder2008classes,ryu2010class,RevModPhys.88.035005},
%described by a winding number or a Chern number for odd and even $d$ respectively, providing the Hamiltonian contains $(d+1)$ anticommuting terms, forming a $(d+1)$D vector space accommodating the $d$D manifold.
A BIS of $m$th-order (m-BIS) can be defined as the zeros of $m$ anticommuting terms, forming a $(d-m)$D manifold embedded in the $d$D BZ.
The pseudospin texture of the rest terms is represented by a $(d+1-m)$D vector, whose winding associated with the m-BIS (of $(d-m)$D) then yields a topological invariant (e.g. the winding number of a 2D vector along a 1D loop) \cite{li2016topological,li2017chiral,zhang2018dynamical,yu2020high}.
The topological invariant defined this way is of genuine appeal, because it can be directly detected through the dynamics by measuring the time-averaged pseudospin texture after a sudden quench, a feat already demonstrated in several  experimental platforms involving ultracold atom systems \cite{sun2018uncover,yi2019observing,song2019observation,wang2020realization}, solid-state
spin systems \cite{wang2019experimental,xin2020experimental,ji2020quantum}, and superconducting circuits \cite{niu2020simulation}.
%The concept of BISs has also been extended to higher orders, which minimize the necessary measurements.

%However, it cannot be directly applied to $Z$-type HOTIs,
%as a higher dimensional vector space is required to support a lower dimensional topological defect \cite{teo2010classes}.
%as a higher dimensional vector space is required to gapped out the lower-order topological boundary states, allowing higher-order ones to emerge in the gap.
Consider now the complex symmetry classes with $n$th-order $Z$-type topology, whose momentum-space Hamiltonian has $(d+n)$ anticommuting terms.
%topological defects have more than one dimension less than the system's dimension,
For $n>1$, the psudospin texture along the vicinity of the above-defined m-BIS yields a $(d+n-m)$-dimensional vector with at least two more degrees of freedom than that of this m-BIS.  As such one can no longer use the simple winding of the psudospin texture associated with this m-BIS to define a topological invariant.
%\red{Note: Ref. \cite{niu2020simulation} is about 2nd-order topological superconductor, which can also be related to normal BISs. However, it is a D class 2nd-order topology described by a $Z_2$ number [also see PRL 123, 177001 (2019)], so shall be not relevant to our cases.}
To characterize $Z$-type HOTIs with BISs, we advocate to use nested BISs,
where a series of BISs denoted as $S_i^m$ are defined for the system, with $m$ the order of each BIS, and $i=1,2,...$ the (hoped) ``nesting" order of these BISs.
Nontrivial higher-order topological phases can be identified if every $S^m_i$ encloses $S^m_{i+1}$, in the same fashion as the Russian nesting doll. The nontrivial HOTI phase is then characterized by the collection of the respective topological invariants associated with these BISs, which are directly accessible via quench dynamics in the same manner as those of conventional BISs.
This nested-BISs approach provides a systematic route towards the engineering of one important category of HOTIs with arbitrary orders of topology in arbitrary dimensions.

 \section{Results}
\subsection{General consideration of nested BISs}
\begin{figure}
\includegraphics[width=1\linewidth]{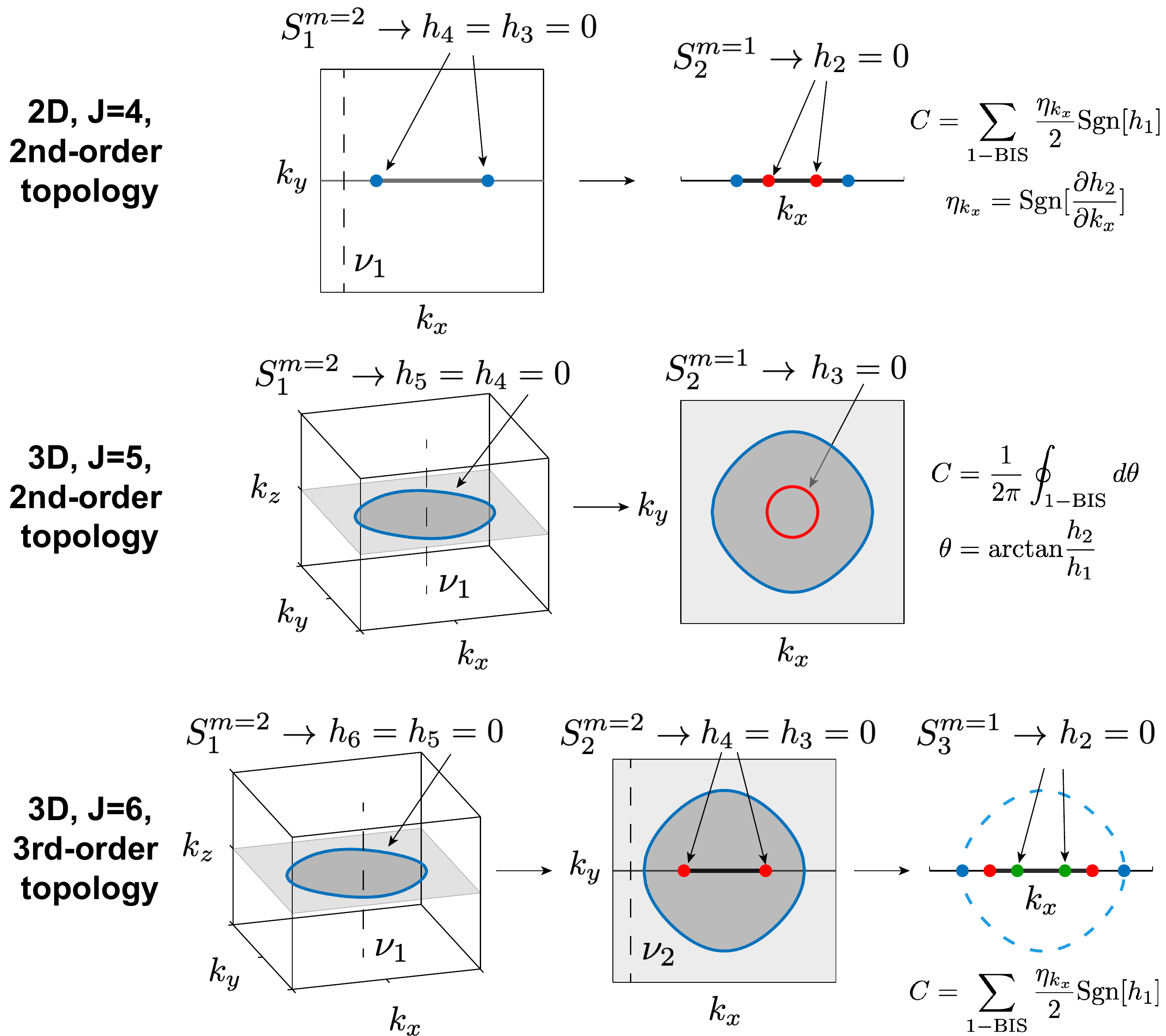}
\caption{Sketches of nested BISs in 2D and 3D. Colored loops and dots represent different BISs ($S_{1,2,3}$), whose conditions are indicated in each panel. In each case, by introducing a 2-BIS of the (sub)system, both the spatial dimension and the order of topology is reduced by {\it one}, from left to right. Each 2-BIS separates regimes with different winding number $\nu_i$ defined for certain terms along the direction perpendicular to the gray lines/planes, i.e., along the black dashed lines. Regimes with nonzero $\nu_i$ are marked with bold lines or darker gray color. The final subsystem of reduction possesses  1st-order topology, characterized by a topological invariant $C$ defined via 1-BIS, as indicated.}
 %\blue{This invariant $C$ is equivalent to a winding number for the 1D BZ of $k_x$ in the first and last rows, and a Chern number for the 2D BZ of $(k_x,k_y)$ in the middle one \cite{li2016topological}.
\label{fig:sketch}
\end{figure}
Consider a general $d$D Hamiltonian with $J$ anticommuting terms,
\begin{eqnarray}
H_d=\mathbf{h}(\mathbf{k})\cdot\mathbf{\gamma}=\sum_{j=1}^J h_j(\mathbf{k})\gamma_j,\label{eq:general}
\end{eqnarray}
with $\gamma_j$ satisfying $\{\gamma_j,\gamma_{j'}\}=2\delta_{jj'}$, $\mathbf{k}=(k_1,k_2,...,k_d)$, and $J=d+n$ is required for the system to host $n$th-order topology (see Supplemental Materials).
Assuming $k_d$ is contained in only the last two terms $h_{J-1}$ and $h_{J}$ (in principle, this condition always arises with some rotation and deformation of the vector space), existence of 1st-order surface states under OBCs along the spatial dimension conjugate with $k_d$ is determined solely by these two terms. Indeed, because the rest terms of the Hamiltonian anticommute with $h_{J-1}$ and $h_{J}$, they preserve the 1st-order surface states and determine their extra features \cite{Mong2011winding,li2017engineering}.
This being the case,  we first define a 2-BIS at the highest nesting level as $S_1^{m=2}$ with  $h_{J}=h_{J-1}=0$, which separates the regimes with and without 1st-order surface states under OBCs along the $d$th dimension, as shown in Fig.~\ref{fig:sketch}. To topologically characterize these regimes, a winding number $\nu_1$ can be defined for a two-component vector $(h_{J-1},h_{J})$ along a 1D trajectory with $k_d$ varying over one period.
{$\nu_1$ takes a quantized value unless this trajectory crosses the 2-BIS, where $(h_{J-1},h_{J})$ vanishes.}
Next, the behavior of the 1st-order surface states thus obtained can be further described by the rest $J-2=(d+n-2)$ terms, forming a $(d-1)$-dimensional subsystem $H_{d-1}$ with a $(n-1)$th-order topology. Furthermore, for the 1st-order surface states to inherit this $(n-1)$th-order topology, a BIS of $H_{d-1}$, which captures its topological information, must fall within the regimes with 1st-order surface states (with nonzero $\nu_1$) bordered by $S_1^{m=2}$ \cite{li2017engineering}.

The above procedure reduces both the order of topology and the system's dimension by {\it one}, and can be repeatedly applied to the subsystem, until a 1st-order topological subsystem finally emerges.  In doing so we should have obtained a series of 2-BIS, $S_i^{m=2}$ with $i=1,2,...n-1$,  each corresponding to the existence of surface states upon open up boundary of one extra direction. One then proceeds to collect a series of winding numbers $\nu_i$ using the respective spin texture operators defining these 2-BISs. The 1st-order topology of the final subsystem thus reached can also be described by its own BIS $S_n^m$, and it is not restricted to $m=2$ because we do not need to further reduce its dimension.

 Fig.~\ref{fig:sketch} illustrates this route of reduction in both dimension and topology.
 %The last-step BIS is 1-BIS,
%\blue{with its topological invariant denoted by $C$ there}.
For the 2D 2nd-order and 3D 3rd-order cases in Fig.~\ref{fig:sketch}, the final subsystem of reduction is a 1D system with two anticommuting terms $h_1$ and $h_2$, possessing 1st-order topology described by a winding number defined from $(h_1,h_2)$ as $k_x$ varies over a period.
In the language of BIS, this winding number can be also obtained by counting the sign of $h_1$ at its 1-BIS of $h_2=0$,
$C=\sum_{\rm 1-BIS}\eta_{k_x}{\rm Sgn}[h_1]/2$.
Here the 1-BIS is given by some discrete points of $k_x$, and $\eta_{k_x}={\rm Sgn}[\partial h_2/\partial k_x]$ represents a sign function yielding the winding direction of the $(h_1,h_2)$ trajectory on these points \cite{yu2020high}.
For such a system to possess higher-order topology, each $S_i^m$ must fall within the regime with a nontrivial $\nu_{i-1}$ determined by $S_{i-1}^{m=2}$, as shown by the colored loops and dots in
Fig.~\ref{fig:sketch}.
The topological properties of the original system are characterized by the winding numbers $\nu_i$ and the 1st-order topological invariant $C$ of the final subsystem.

%Assuming $k_d$ is contained in no more the last $n$ terms of $(h_{J-n+1},h_{J-n+2},...,h_{J})$ (for general system, this condition may be obtained with some rotation and deformation of the vector space), a $n$-BIS can be defined for them and the pseudospin texture of the rest is reduced to a $(d-1)$D system with $J-n$ terms. To reduce the order of the topology, we also need $n>1$ for defining first $n$-BIS \blue{[Cite or explain in supplementary material? Can be seen simply from constructing anticommuting matrices from multiple sets of Pauli Matrices.]}. Here we choose $n=J-d$

\subsection{3D 2nd-order topological insulators}
\begin{figure*}
\includegraphics[width=1\linewidth]{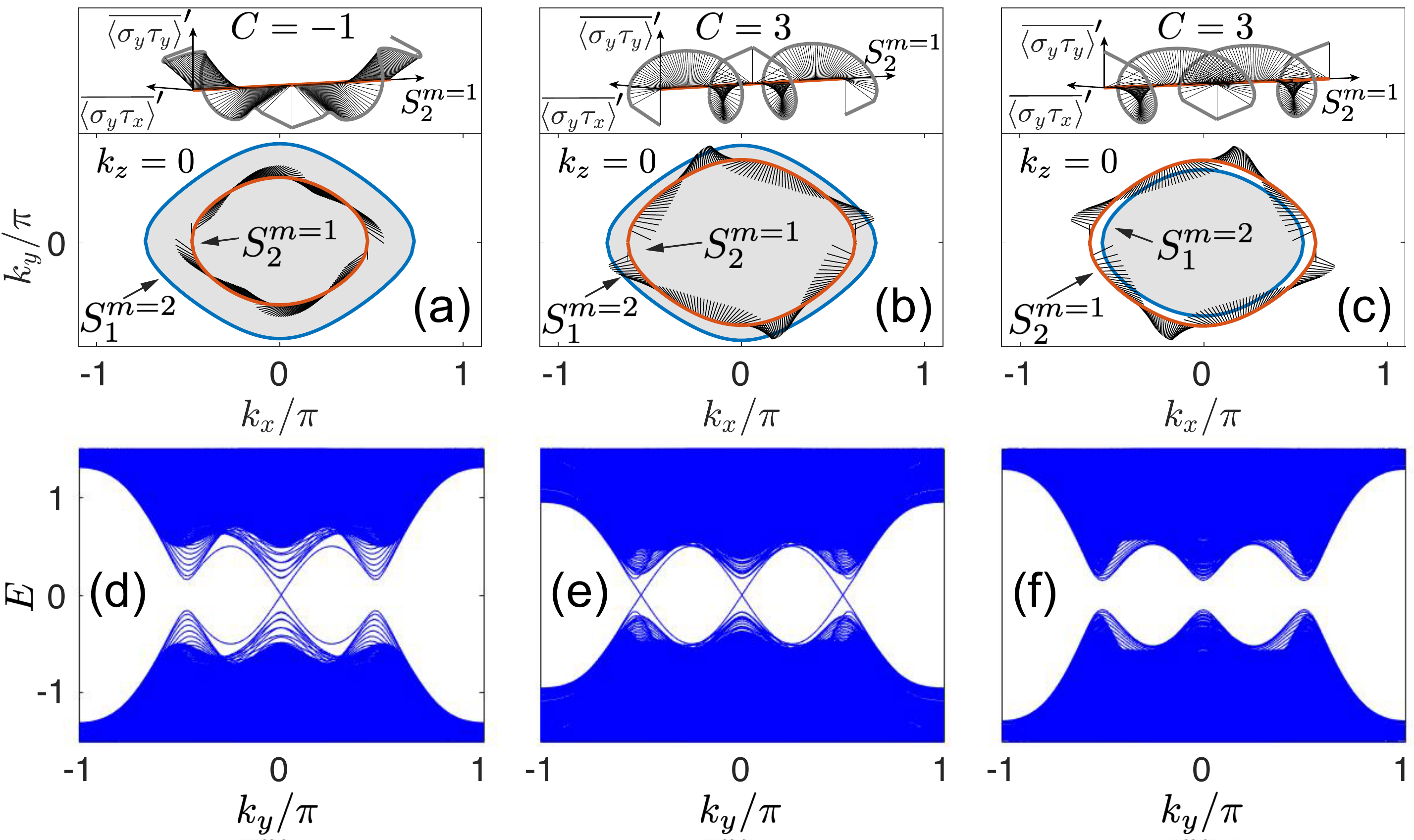}
\caption{A 3D 2nd-order topological insulator.
(a-c) BISs of a 3D system with 2nd-order topology with (a) $m_0=1.5$, $m_1=1.2$; (b) $m_0=1.5$, $m_1=0.8$; and (c) $m_0=2$, $m_1=0.8$ respectively. Other parameters are $\lambda=0.5$, and $t_0=t_1=t_z=1$.
(d-f) $xz$-OBC spectra corresponding to (a-c). The in-gap states are two-fold degenerate, localized at the hinges of top and bottom surfaces along $z$ direction.
In (a-c), blue loops are the 2-BIS $S_{1}^{m=2}$ given by $h_{x0}=h_{z0}=0$, which separates $k_x-k_y$ plane into regimes with $\nu_1=1$ (shadowed) and $\nu_1=0$ (white).
Red loops are the 1-BIS $S_{2}^{m=1}$ of the subsystem given by $h_{xz}=0$, and the black short lines ``growing" from them indicate the normalized time-averaged pseudospin texture
% $\mathcal{N}(\overline{\langle\sigma_y\tau_x\rangle},\overline{\langle\sigma_y\tau_y\rangle})$
in quantum quench dynamics along $S_{2}^{m=1}$.
Top panels of (a-c) illustrate the winding of the time-averaged pseudospin texture along $S_{2}^{m=1}$. In (c,f), $S_{2}^{m=1}$ falls in the regime with $\nu_1=0$, thus the system is topologically trivial even though it has a winding number of $C=3$ along $S_{2}^{m=1}$.}
\label{fig:3D_2nd}
\end{figure*}

As a concrete example, we now consider a 3D system with 2nd-order topology described by Eq.~(\ref{eq:general}) with $J=5$. The five anticommunting $\gamma_j$ matrices can be constructed by two sets of Pauli matrices $\sigma$, $\tau$ and the identity matrix. Without further restriction of each term of $\gamma_j$, the system below belongs to the A class with no additional symmetry, and may support topological boundary states along odd-dimensional boundaries \cite{teo2010classes}.
The explicit system we consider is described by the Hamiltonian $H(\mathbf{k})=H_1(\mathbf{k})+H_2(\mathbf{k}_{\parallel})$,
with
\begin{eqnarray}
H_1(\mathbf{k})&=&\left[t_0(\cos k_x+\cos k_y+\cos k_z)-m_0\right]\sigma_x\tau_0\nonumber\\
&&+t_z\sin k_z\sigma_z\tau_0,\\
H_2(\mathbf{k}_{\parallel})&=&\lambda \sin 2k_y\sigma_y\tau_x-\lambda\sin 2k_x\sigma_y\tau_y\nonumber\\
&&+\left[m_1-t_1(\cos k_x+\cos k_y)\right]\sigma_y\tau_z.
\end{eqnarray}
Here $\mathbf{k}_{\parallel}=(k_x,k_y)$ and $k_z$ is contained only in $H_1$, allowing us to reduce both the spatial dimension and the order of topology by defining the $S_1^{m=2}$ with $H_1=0$.  $\tau_a$ and $\sigma_a$ with $a=x,y,z$ are two sets of Pauli matrices, and $\tau_0$ is a $2\times2$ identity matrix acting in the subspace of $\tau_a$.
In the following discussion, we denote the coefficient of $\sigma_a\tau_{a'}$ as $h_{aa'}$. Upon taking OBCs along $z$ direction, existence of 1st-order boundary states at each $\mathbf{k}_{\parallel}$ can be characterized by a winding number defined as
\begin{eqnarray}
\nu_1(\mathbf{k}_{\parallel})=\frac{1}{2\pi}\oint_{k_z}\frac{h_{x0}dh_{z0}-h_{z0}dh_{x0}}{h_{x0}^2+h_{z0}^2}.
\end{eqnarray}
As shown in Fig. \ref{fig:3D_2nd}(a-c), the $k_x-k_y$ plane is divided into regimes with different values of $\nu_1$, the boundary between which has vanishing $(h_{x0},h_{z0})$ at $k_z=0$, forming a 2-BIS $S_1^{m=2}$ of the system.
The behavior of the 1st-order boundary states within the regime of nonzero $\nu_1$ can be described by an effective 2D Hamiltonian of $H_2(\mathbf{k}_{\parallel})$, which has its own 1st-order topology characterized by a winding of $(h_{yx},h_{yy})$ along its (nested) 1-BIS $S_2^{m=1}$ given by $h_{yz}=0$.

The \blue{BISs and the} winding behavior of the pseudospin texture associated with each \blue{of them} can be observed by time-averaged pseudospin textures in quantum quench dynamics \cite{zhang2018dynamical}.
\blue{Nevertheless, different initial states are required for determining the zeros (which give the BISs) and the winding behavior of the pseudospin texture.
For a concerned pseudospin component $\sigma_a\tau_{a'}$,
we shall consider two of its time-averaged values in quantum quench dynamics with different initial states, denoted as $\overline{\langle\sigma_a\tau_{a'}\rangle}$ and $\overline{\langle\sigma_a\tau_{a'}\rangle}'$ respectively. The first one gives the BISs with its zeros, and the latter one gives the topological invariants with its winding behavior.}
Additional details are presented in the Supplemental Materials.

As shown in Fig.~\ref{fig:3D_2nd},  the number of gapless hinge states under $xz$-OBC has a one-to-one correspondence with the winding number $C$ obtained from the time-averaged pseudospin texture, provided that $S_2^{m=1}$ falls within the nontrivial regime with $\nu_1=1$ enclosed by $S_1^{m=2}$.  Remarkably, if $S_2^{m=1}$ falls outside this nontrivial regime [Figs. \ref{fig:3D_2nd}(c,f)], there is no gapless hinge state within the bulk gap even when the time-averaged pseudospin texture assumes a nonzero value, thus confirming the importance of having nested-BISs towards characterization of HOTIs. Note that in general $\nu_1$ may take values other than 1 or 0, then the total number of hinge states is given by $\nu_1\times C$ if the BISs are indeed nested (see Supplemental Materials).
%Because $\nu_1$ in the explicit model here is either $1$ or $0$, thus the number of hinge states is either equal to the %winding number $C$ as in Fig.~\ref{fig:3D_2nd}(a,b,d,e), or zero as in Fig.~\ref{fig:3D_2nd}(c,f).
It is now also fruitful to investigate the consequences of the crossing
of two particular BISs as we tune the system parameters.  For example, due to this crossing the number of gapless hinge states may only partially reflect the winding number $C$ of the spin texture associated with $S_2^{m=1}$.  More analysis is presented in Supplemental Materials.

%This scenario can be understood through defining a 2-BIS $S_2^{m=2}$ of the subsystem and analyzing its geometrical relation to $S_1^{m=2}$ of the parent system, as discussed in Supplemental Materials \cite{SuppMat}. }

\subsection{3D 3rd-order topological insulators}
\begin{figure}
\includegraphics[width=1\linewidth]{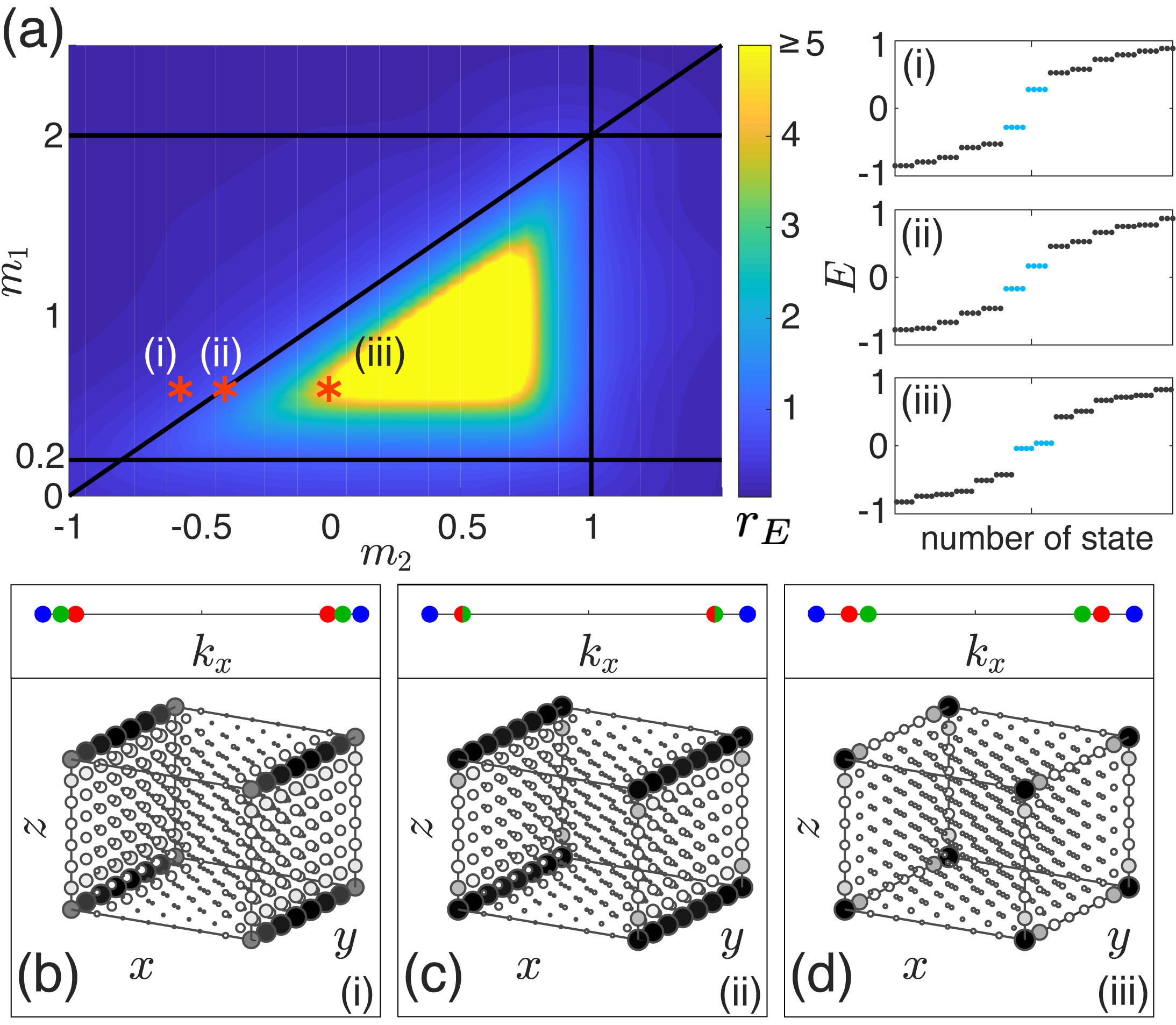}
\caption{A 3D 3rd-order topological insulator.
(a) A phase diagram of the 3rd-order topological insulators. The color bar indicates the energy-spacing ratio $r_E$.
Black lines are the topological phase boundaries corresponding to Eqs. (\ref{eq:condition1}) and (\ref{eq:condition2}). The three insets on the right illustrate the eigenenergies close to $E=0$ for the three red stars in the diagram.
(b)-(d) BISs at $k_y=k_z=0$ and the summed spatial distribution $\rho(x,y,z)$ of the middle eight states marked with cyan in the spectrum, corresponding to the three red stars in (a) respectively.
The blue, red, and green dots on the top panels are the BISs of $S_1^{m=2}$, $S_2^{m=2}$, and $S_3^{m=1}$, given by $H_1=0$, $H_2=0$, and $h_{yyx}=0$ respectively.
The winding number $\nu_1$/$\nu_2$ takes nonzero value within the regime bordered by blue/red dots.
The other parameter is set to $m_0=1.2$.
The system's size is $8\times8\times8$, with eight sublattices in each unit cells.
}
\label{fig:3rd_order}
\end{figure}
To further demonstrate how the use of nested BISs offers a powerful scheme to investigate and engineer HOTIs, next we consider a class of 3D 3rd-order topological insulators depicted by Eq.~(\ref{eq:general}) with $J=6$.
As an example, we assume $H=H_1(\mathbf{k})+H_2(\mathbf{k}_\parallel)+H_3(k_x)$ with
\begin{eqnarray}
H_1(\mathbf{k})=h_{x00}\sigma_x\tau_0{s}_0+h_{z00}\sigma_z\tau_0{s}_0,\nonumber\\
H_2(\mathbf{k}_\parallel)=h_{yx0}\sigma_y\tau_x{s}_0+h_{yz0}\sigma_y\tau_z{s}_0,\nonumber\\
H_3(k_x)=h_{yyx}\sigma_y\tau_y{s}_x+h_{yyz}\sigma_y\tau_y{s}_z,
\end{eqnarray}
$(\sigma_a,\tau_a,{s}_a)$ three sets of Pauli matrices for $a=x,y,z$, and corresponding $2\times2$ identity matrices for $a=0$. The Hamiltonian is again a sum of operators from a Clifford algebra.
This Hamiltonian satisfies the chiral symmetry $SHS^{-1}=-H$ with $S=\sigma_y\tau_y{s}_y$, thus it belongs to the AIII class and may possess even-dimensional topological boundary states (such as 0D corner states).
In our considerations of nested BISs we do not need assistance from this chiral symmetry.
To be more explicit we consider
\begin{eqnarray}
h_{z00}=\sin k_z,&~&h_{x00}=\cos k_x+\cos k_y+\cos k_z-m_0,\nonumber\\
h_{yz0}=\sin k_y,&~&h_{yx0}=\cos k_x+\cos k_y-m_1,\nonumber\\
h_{yyz}=\sin k_x,&~&h_{yyx}=\cos k_x-m_2,
\end{eqnarray}
so that $z$ direction is reduced with the first 2-BIS $S_1^{m=2}$ as zeros of $H_1$, $y$ direction is reduced by the second 2-BIS $S_2^{m=2}$ as zeros of $H_2$, and $H_3$ describes a 1D subsystem with 1st-order topology, whose 1-BIS $S_3^{m=1}$ can be defined as the points with $h_{yyx}=0$. The 3rd-order topological properties here are thus  featured by three topology invariants, i.e. two winding numbers $\nu_1(\mathbf{k}_\parallel)$ and $\nu_2(k_x)$ found from $H_1$ and $H_2$ respectively, and a topological invariant $C$ corresponding to the sign of $h_{yyz}$ on the 1-BIS of $h_{yyx}=0$. In this system, $C$ is also equivalent to another winding number defined from $H_3$ with $k_x$ varying over one period.
Following these definitions, nontrivial 3rd-order topology of this system is found to require
\begin{eqnarray}
m_0<3,~m_1<2,~m_2<1,\label{eq:condition1}
\end{eqnarray}
%$m_0<3$, $m_1<2$, and $m_2<1$,
so that each topological invariant takes nonzero values in certain regions in the BZ; and
\begin{eqnarray}
m_2>m_1-1>m_0-2,\label{eq:condition2}
\end{eqnarray}
%$m_2>m_1-1>m_0-2$,
so that each BIS $S^m_{i}$ falls within the nontrivial regime determined by $S^m_{i-1}$.
%is nested in that of its parent (sub)system.

 \begin{figure*}
\includegraphics[width=1\linewidth]{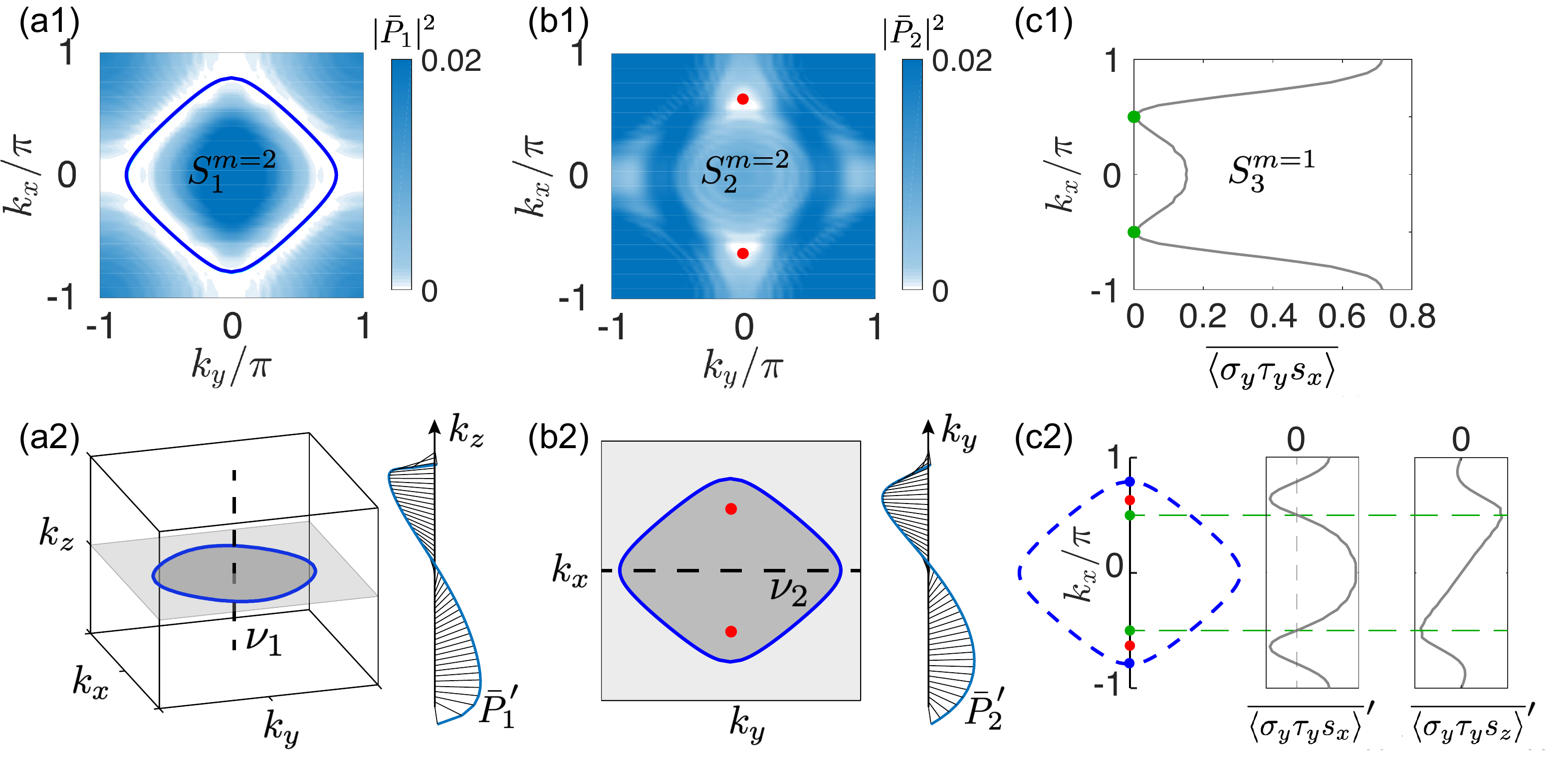}
\caption{Dynamical chracterization of a 3D 3rd-order topological insulator.
(a1) Time-averaged pseudospin amplitude $|\bar{P}_1|^2$, whose zeros gives the first BIS $S^{m=2}_1$ (blue loops). As $\bar{P}_1$ is a two-component vector, $S^{m=2}_1$ is a 1D loop in the 3D BZ, and we have only shown the 2D $k_x-k_y$ plane containing this loop.
(a2) $S^{m=2}_1$ in the 3D BZ, and the winding of the time-averaged pseudospin vector $\bar{P}_1'$ along the dash line, corresponding to a winding number $\nu_1(\mathbf{k}_\parallel)=1$ in the regime enclosed by $S^{m=2}_1$.
(b1) Time-averaged pseudospin amplitude $|\bar{P}_2|^2$ associated with the second BIS $S^{m=2}_2$ (red dots). $\bar{P}_2$ is a two-component vector containing only $k_x$ and $k_y$, therefore $S^{m=2}_1$ are given by some points in the 2D BZ.
(b2) $S^{m=1}_2$ and $S^{m=2}_2$ in the 2D BZ of $k_x$ and $k_y$, and the winding of the time-averaged pseudospin vector $\bar{P}_2'$ along the dash line, corresponding to a winding number $\nu_2(k_x)=1$ in the regime enclosed by $S^{m=2}_2$.
(c1) Time-averaged pseudospin texture $\overline{\langle\sigma_y\tau_y s_x\rangle}$ associated with the third BIS $S^{m=1}_3$ (green dots).
\blue{(c2) The three BISs in the 1D BZ of $k_x$, and the time-averaged pseudospin texture of $\overline{\langle\sigma_y\tau_y s_x\rangle}'$ and $\overline{\langle\sigma_y\tau_y s_z\rangle}'$.
A nonzero topological invariant $C=-1$  can be obtained as both of $\overline{\langle\sigma_y\tau_y s_z\rangle}'$ and the derivative of $\overline{\langle\sigma_y\tau_y s_x\rangle}'$ take opposite signs at the two points of $S^{m=1}_{3}$ (green dots).
Notations with and without an apostrophe correspond to quantities obtained with different pre-quench Hamiltonians, as elaborated in the Supplemental Materials.}
Parameters are $m_0=1.2$, $m_1=0.6$, and $m_2=0$.
}
\label{fig:dynamic_3rd}
\end{figure*}

Fig.~\ref{fig:3rd_order}(a) illustrates a phase diagram of the system, where the phase boundaries (black lines) are given by Eqs. (\ref{eq:condition1}) and (\ref{eq:condition2}) with the inequality signs replaced by equal signs. The topologically nontrivial phase is given by the triangle regime in the center.
The three insets \GJB{on} the right illustrate the OBC spectra \GJB{associated with parameter choices indicated by the three red stars labeled in Fig.~\ref{fig:3rd_order}(a).  The shown spectra} are four-fold degenerate due to the extra two sets of Pauli matrices.
Eight nearly degenerate in-gap states are seen in the nontrivial regime, which merge into the bulk spectrum when decreasing $m_2$.
The BISs at $k_y=k_z=0$ and  the summed spatial distribution $\rho(x,y,z)=\sum_{\beta,\alpha}|\psi_{\beta,\alpha}(x,y,z)|^2$ of the middle eight states for the three chosen points are shown in Fig. \ref{fig:3rd_order}(b)-(d), with $\psi_{\beta,\alpha}(x,y,z)$ the amplitude of the $\beta$th in-gap eigenstate at the unit cell located at $(x,y,z)$, and $\alpha$ the sublattice index.
A 3rd-order topological phase transition occurs when $S_3^{m=1}$ overlaps with $S_2^{m=2}$ at a phase boundary [Fig. \ref{fig:3rd_order}(c)],
and eight in-gap states localized at eight corners emerge when the system enters a nontrivial 3rd-order topological phase [Fig. \ref{fig:3rd_order}(d)].
%A detailed study of the dynamical characterization of the three cases is elaborated in Supplementary Materials \cite{SuppMat}.
%In the left four columns of Fig.~\ref{fig:3rd_order}, we illustrate the BISs at $k_y=k_z=0$, $xyz$-OBC spectra, and the summed spatial distribution $\rho(x,y,z)=\sum_{\beta,\alpha}|\psi_{\beta,\alpha}(x,y,z)|^2$ of the middle eight in-gap states, with $\psi_{\beta,\alpha}(x,y,z)$ the amplitude of the $\beta$th in-gap eigenstate at the unit cell located at $(x,y,z)$, and $\alpha$ the sublattice index.
%We can also see that all eigenenergies are four-fold degenerate, a result from the extra two sets of Pauli matrices.
%A 3rd-order topological phase transition with decreasing $m_2$ is demonstrated from left to right.  For $m_2=0>m_1-1$,  eight in-gap states localized at eight corners are seen.
%As $m_2$ decreases and becomes smaller than $m_1-1$, these states become delocalized along hinges, and merge into the bulk spectrum.
Note that due to the considered small system size  ($8$ unit cells along each direction), the gap closing at the phase transition point is not so clear in Fig.~\ref{fig:3rd_order}(a), with the degeneracy of the in-gap states slightly lifted even in the nontrivial regime. To further verify the topological phase boundaries,
we next calculate an energy-spacing ratio
\begin{eqnarray}
r_E=\frac{E_{+,1}-E_{+,0}}{E_{+,0}-E_{-,0}},
\end{eqnarray}
with $E_{\pm,0}$ the eigenenergies of the positive and negative branches of the eight in-gap states respectively, and $E_{+,1}$ the lowest positive eigenenergy of the rest eigenstates, all computationally obtained in our finite system with 8 unit cells only. In a topological nontrivial regime, $r_E$ shall take a large value as the topological in-gap states are almost degenerate. On the other hand, $r_E$ is expected to decrease rapidly when the system moves into a topological trivial regime, where the middle eight states also belong to the bulk spectrum. In Fig. \ref{fig:3rd_order}(a) we have presented $r_E$ as a function of $m_1$ and $m_2$, and the results agree very well with the topological phase boundaries obtained from the BISs.

%To dynamically characterize the 3rd-order topological properties of this model, we consider two steps of dynamical characterization,
The 3rd-order topological properties of this model can also be characterized through quench dynamics,
 as \GJB{elaborated} in the Supplemental Materials.
\GJB{In short}, the three BISs $S_1^{m=2}$, $S_2^{m=2}$, and $S_3^{m=1}$ are given by the zeros of the \GJB{time-averaged} pseudospin vector
\begin{eqnarray}
\bar{P}_1=(\overline{\langle\sigma_z\tau_{0} s_{0}\rangle},\overline{\langle\sigma_x\tau_{0} s_{0}\rangle}),\nonumber\\
\bar{P}_2=(\overline{\langle\sigma_y\tau_{z} s_{0}\rangle},\overline{\langle\sigma_y\tau_{x} s_{0}\rangle}),\nonumber
\end{eqnarray}
and $\overline{\langle\sigma_y\tau_{y} s_{x}\rangle}$, respectively, as shown in Fig.~\ref{fig:dynamic_3rd}(a1)-(c1).
Among them, $S_1^{m=2}$ is a 1D loop in the 3D BZ.  Indeed, A line solution is obtained when requiring the two components of $\bar{P}_1$ (as functions of the 3D momentum $\mathbf{k}=(k_x,k_y,k_z)$) to be zero, and the projection of this line on the $(k_x, k_y)$ plane is plotted in Fig.~\ref{fig:dynamic_3rd}(a1) as $S_1^{m=2}$.   $\bar{P}_2$ also has two components as functions of the 2D momentum $\mathbf{k}_\parallel=(k_x,k_y)$. Requiring both components of $\bar{P}_2$ to vanish yields  $S_2^{m=2}$ as some points in the shown 2D BZ.  Likewise, $S_3^{m=1}$ is formed by some points in the 1D BZ of $k_x$ as $\overline{\langle\sigma_y\tau_{y} s_{x}\rangle}$ is a scalar depending only on $k_x$.
As shown in Fig.~\ref{fig:dynamic_3rd}(a2), (b2), by further measuring the time-averaged pseudospin texture of different pseudospin components along $k_z$ and $k_y$, one can see that the first winding number $\nu_1(\mathbf{k}_\parallel)=1$ for $\mathbf{k}_\parallel$ within $S_1^{m=2}$, and the second winding number $\nu_1(k_x)=1$ for $k_x$ between the two points of $S_2^{m=2}$.
\blue{As elaborated in Supplemental Materials, to actually obtain these topological invariants from the dynamics, one may consider quantities $\bar{P}'_{1,2}$ instead, with the only difference between $\bar{P}_{1,2}$ and $\bar{P}_{1,2}'$ being that they are obtained from different pre-quench Hamiltonians. These considerations can ensure that $\bar{P}_{1,2}$ vanishes only at zeros of the concerned pseudospin vectors and gives the desired BISs, and furthermore $\bar{P}_{1,2}'$ are able to capture the winding of these vectors.}

The third topological invariant of the subsystem $H_3(k_x)$, following our discussion of Fig. \ref{fig:sketch}, is defined as
\begin{eqnarray}
C=\sum_{S_3^{m=1}}\frac{1}{2}{\rm Sgn}[h_{yyz} \frac{\partial h_{yyx}}{\partial k_x}].
\end{eqnarray}
From the above-mentioned quench dynamics, $h_{yyx}$ and $h_{yyz}$ in the above expression can be effectively captured by the time-averaged pseudospin texture $\overline{\langle\sigma_y\tau_{y} s_{x}\rangle}'$ and $\overline{\langle\sigma_y\tau_{y} s_{z}\rangle}'$ illustrated in Fig.~\ref{fig:dynamic_3rd}(c2).
\blue{Similar to the other two winding numbers, here $\overline{\langle\sigma_y\tau_{y} s_{x}\rangle}'$ is also distinguished from $\overline{\langle\sigma_y\tau_{y} s_{x}\rangle}$ in Fig.~\ref{fig:dynamic_3rd}(c1), as they correspond to different pre-quench Hamiltonians (see Supplemental Materials).}
Thus the topological invariant $C$ can be expressed as
\begin{eqnarray}
C=\sum_{S_3^{m=1}}\frac{1}{2}{\rm Sgn}[\overline{\langle\sigma_y\tau_{y} s_{z}\rangle}' \frac{\partial \overline{\langle\sigma_y\tau_{y} s_{x}\rangle}'}{\partial k_x}].
\end{eqnarray}
From Fig.~\ref{fig:dynamic_3rd}(c2), we can read out from the slopes and the values of the plotted quantities that $C=-1$ for all the three presented cases.
%Nevertheless, only in the first row the three BISs are nested in the correct order (each $S_{i}^m$ encloses $S_{i+1}^m$), leading to the topologically nontrivial case in Fig. 3(d) in the main text.
%The second and third rows of Fig. \ref{fig:dynamic_3rd} respectively correspond to the phase transition point and the trivial insulating phase in Fig.~3(c) and Fig.~3(b) of the main text.
Note that here we have chosen the topologically nontrivial case of Fig. \ref{fig:3rd_order}(d). For the other two cases of Fig. \ref{fig:3rd_order}(b) and Fig. \ref{fig:3rd_order}(c), the obtained time-averaged pseudospin texture is similar to that of Fig.~\ref{fig:dynamic_3rd}, except that $S_3^{m=1}$ falls outside or overlaps $S_2^{m=2}$, indicating a trivial insulating phase or a phase transition point.

\section{Discussion}
We have shown that  one important class of HOTIs with arbitrary orders of topology in arbitrary dimensions can be characterized and {\it dynamically detected} by considering nested BISs. Comparing with the nested Wilson loops treatment, there are three important features inherent in our approach. Firstly, the BISs at different nesting levels can be treated under equal footing, with their geometrical relations becoming a key insight to digest higher-order topological phase transitions. Secondly,  the entire collection of topological invariants based on BISs are measurable in ongoing experiments by quantum quench dynamics, just as how they are measured in probing 1st-order topology.  Thus, topological invariants at each and every order of topology can be dynamically characterized. Thirdly, our topological characterization does not require any additional crystal symmetry to facilitate the investigations.
Among the several experimental platforms where BISs have been examined,
nested BISs can be dynamically detected in solid-state qubit systems \cite{wang2019experimental,xin2020experimental,ji2020quantum,niu2020simulation}, where momentum space can be simulated by the highly tunable parameter space of certain qubit Hamiltonians.
In ultracold atom systems, time-averaged pseudospin textures can be also readily measured with time-of-flight imaging \cite{sun2018uncover,yi2019observing,song2019observation,wang2020realization}. Hence physical insights based on nested BISs are expected to be very useful in guiding and then dynamically verifying the engineering of HOTIs with designed optical lattice potentials.

 %In practice, the corresponding topological invariants can be detected by measuring the time-average %pseudospin texture in certain regimes in the BZ after a sudden quench, which provides an efficient scheme %for experimental detecting the topological invariants of HOTIs, even without the assistance of crystal %symmetries.
%The nested BISs also unveil a natural comprehension of higher-order topology as topology of not only the %bulk, but also the boundary of the system.

%Indeed, the only assumption we made is that the momentum-space Hamiltonian is expressible in terms of operators from  (part of) a Clifford algebra.
%But even this requirement can be relaxed case by case. For example, in the 3D 2nd-order topological system considered here,  one may introduce an extra $\sigma_y\tau_0$ term, which anticommutes with $H_1(\mathbf{k})$ that gives $S_1^{m=2}$, but commutes with $H_2(\mathbf{k}_\parallel)$. This term only modifies the eigenenergies, without changing the topology of $H_2(\mathbf{k}_\parallel)$ described by the invariant $C$, or the existence of 1st-order boundary states determined by the invariant $\nu_1$ obtained from $H_1(\mathbf{k})$.  Hence our approach still applies.  Certainly even more investigations are needed to further extend our approach.

% However, the extension of our method is still not clear for systems with more complicated commuting and %anticommuting relations of the Hamiltonian, which awaits further investigations.

\bigskip

\begin{acknowledgements}
{\bf Acknowledgements:}
J.G. acknowledges fund support by the Singapore Ministry of Education Academic
Research Fund Tier-3 Grant No. MOE2017-T3-1-001
(WBS. No. R-144-000-425-592) and by the Singapore National Research Foundation
Grant No. NRF-NRFI2017- 04 (WBS No. R-144-000-378-
281).
\end{acknowledgements}

\onecolumngrid
\begin{center}
\textbf{\large Supplementary Materials}\end{center}
\setcounter{equation}{0}
\setcounter{figure}{0}
\renewcommand{\theequation}{S\arabic{equation}}
\renewcommand{\thefigure}{S\arabic{figure}}
\renewcommand{\cite}[1]{\citep{#1}}

\section{Comparison between the nested BISs and several different topological characterizations of HOTIs}
Since the discovery of HOTIs, various methods of their topological characterization have been proposed for different systems with different types of topology, as briefly summarized in Table \ref{table1}. The earliest one is the method of ``nested" Wilson loops \cite{benalcazar2017HOTI,Benalcazar2017HOTI2}, which are usually used to describe quantized electric multipole insulators. In 2D and 3D systems, this method allows topological characterizations of corner modes and hinge modes with the winding number or Chern number of the Wannier bands. More generally, the nested Wilson loops map a HOTI to a lower-order topology of its Wannier bands. In principle, we can always reduce an arbitrary HOTI to a 1st-order topological system by repeatedly \GJ{applying} this method. 
However, the topological characterization of the final 1st-order topological system and its experimental detection still require case-by-case investigations.

The symmetry indicator approach \cite{fu2007,bradlyn2017topological,po2017symmetry,ono2020refined,tang2019comprehensive,vergniory2019complete,zhang2019catalogue} has been widely used in characterizing conventional (1st-order) topological phases with $\mathbf{k}$-dependent symmetries, and also in describing HOTIs with fractional charge \GJ{polarization}  \cite{Benalcazar2017HOTI2,ezawa2018HOTI,benalcazar2019quantization}. 
\GJ{In these cases}  a transition of topology always occurs in pairs of momenta mapped to each other by a specific symmetry, which acquires opposite topological charges after the transition, resulting in unchanged topological properties of the systems. At high-symmetric points, however, the momentum is mapped to itself by the symmetry, so that a nonzero topological charge can be acquired by the system. Therefore for a HOTI, topological properties such as polarization or fractional charge of the system can be determined solely by information of eigenfunction at some high-symmetric points, providing a convenient method in the presence of certain symmetries.

The 2D Zak phase $\mathbf{P}=(\gamma_x,\gamma_y)$ was first proposed for describing a 1st-order topology in 2D systems with zero Berry curvature \cite{liu2017novel}, 
defined as
\begin{eqnarray}
\gamma_\alpha=\frac{1}{2\pi}\int_{\rm BZ}dk_x dk_y {\rm Tr}[A_\alpha(k_x,k_y)],
\end{eqnarray}
with $A_\alpha=\langle\psi|i\partial_\alpha|\psi\rangle$ the Berry connection.
This method has been found to also have a correspondence with corner states in the presence of mirror symmetries \cite{xie2018second}.
In such systems, the 2D Zak phase describes the polarizations along each direction, and can be determined by the symmetry indicator associated with the mirror symmetries.
On the other hand, from the definition, $\gamma_\alpha$ is given by first calculating a 1D Zak phase defined along $\alpha$-direction, then taking average  over all $k_{\alpha'}$, with $\alpha\neq\alpha'$.
It is also known that a quantized Zak phase is equivalent to a winding number and can also be protected by a chiral or $PT$ symmetry, suggesting that the quantized 2D Zak phase may also be extendable to systems without mirror symmetries.

The boundary winding number is proposed for specific models constructed by different sets of Su-Schrieffer-Heeger (SSH) models along different directions \cite{li2018direct,serra2019observation}. This method requires no crystal symmetry either, and can predict various configurations of corner states in systems with different spatial dimensions \cite{li2018direct}. Experimentally, this winding number can be directly measured in LC circuit lattice \cite{serra2019observation}.
Nevertheless, this method has only been shown to be applicable to high-dimensional extensions of the SSH model.
% where each couplings is restricted to only one spatial dimension.

The diagonal winding number has been used to describe topological quadrupole insulator \cite{imhof2018topolectrical} and
Bernevig-Hughes-Zhang (BHZ) model \cite{bernevig2006quantum} with additional mass terms \cite{seshadri2019generating}. 
This topological invariant is defined along the diagonal lines $k_x=\pm k_y$ in the 2D BZ of the studied systems, where several anticommuting terms or their linear combinations vanish (usually due to some symmetries of the system), resulting in a 2-component Hamiltonian vector possessing a quantized winding along the 1D diagonal lines.
These diagonal lines can also be comprehended as a BIS of the system with certain (mirror) symmetries, leading to an alternative method to characterize higher-order topology through BISs. However, it is unclear to what extent this method can be generalized to other models.

Finally, the ``nested" BISs proposed in this paper requires that the Hamiltonian vector containing only anticommuting terms, i.e., \GJ{a sum of operators as part of a Clifford algebra}. 
This condition represents one category of $Z$-type HOTIs, with no restriction of the order of topology or the system dimension. Furthermore, the topological invariants related to the BISs can be detected by measuring the time-averaged pseudospin texture after quenching a pseudospin polarized system to a topologically nontrivial one.  In this sense, the nested BISs approach proposed in this work is highly useful for experimental studies.
On the other hand, the requirement of Clifford algebra can also be relaxed case by case. For example, in the 3D 2nd-order topological system considered here,  one may introduce an extra $\sigma_y\tau_0$ term, which anticommutes with $H_1(\mathbf{k})$ that gives $S_1^{m=2}$, but commutes with $H_2(\mathbf{k}_\parallel)$. This term only modifies the eigenenergies, without changing the topology of $H_2(\mathbf{k}_\parallel)$ described by the invariant $C$, or the existence of 1st-order boundary states determined by the invariant $\nu_1$ obtained from $H_1(\mathbf{k})$.  Hence our approach still applies.  Certainly even more investigations are needed to further extend our approach.

\begin{table}[h]
\footnotesize
\renewcommand{\arraystretch}{1.5}
%\begin{tabular}{|l|l|l|l|l|l|l|}\hline
%\resizebox{\linewidth}{60pt}
{
\begin{tabular}{|c|c|c|}
\hline  
\bf{Method} & \tabincell{c}{\bf{Applicable} \bf{topology}} & \bf{Requirements}  \\
\hline 
\tabincell{c}{``nested" Wilson loops} \cite{benalcazar2017HOTI,Benalcazar2017HOTI2} & arbitrary & \tabincell{c}{topology of Wannier bands}  \\
\hline  
\tabincell{c}{symmetry indicator} \cite{Benalcazar2017HOTI2,ezawa2018HOTI,benalcazar2019quantization} & arbitrary & $\mathbf{k}$-dependent symmetries\\
\hline  
2D Zak phase \cite{liu2017novel,xie2018second} & \tabincell{c}{2nd-order topology in 2D} & \tabincell{c}{mirror symmetry along each direction} \\
\hline 
\tabincell{c}{boundary winding numbers} \cite{li2018direct,serra2019observation}& \tabincell{c}{$d$nd-order topology in $d$D} & \tabincell{c}{extended $d$D SSH model} \\
\hline 
\tabincell{c}{diagonal winding numbers} \cite{imhof2018topolectrical,seshadri2019generating} & \tabincell{c}{2nd-order topology in 2D} & 2-component Hamiltonian vector along diagonal lines in the BZ \\
\hline 
\tabincell{c}{``nested" BISs (this work)}& arbitrary & \tabincell{c}{Clifford algebra}\\
\hline 
\end{tabular}\smallskip
}%\\   1. equivalent to a lower-order topology of Wannier bands.
%\\    2. extendable to corner states in higher dimensions.
%\captionsetup{justification=centerlast}
\caption{Comparison between a few different topological characterizations of HOTIs. 
%The ``nested" Wilson loops map a HOTI to a lower-order one of its Wannier bands, thus a clear lower-order topological description is required. The corresponding topological invariant may be difficult to detected for some cases, e.g. a high-dimension winding number or Chern number for the Wannier bands exhibiting a 1st-order high-dimensional topology.
%The 2D Zak phase is specifically defined for describing corner states in 2D systems with 1st-order topology along each direction, but its definition can be directly extended to systems in higher dimensions. The boundary winding numbers also corresponds to the existence of corner states, providing each boundary of the system is analogous to a Su-Schrieffer-Heeger (SSH) model \cite{SSH}. The diagonal winding number is defined for a variant of the Bernevig-Hughes-Zhang (BHZ) model \cite{seshadri2019generating,bernevig2006quantum}.
%The ``nested" BISs proposed in this paper requires that the Hamiltonian vector containing only anticommuting terms, satisfying the Clifford algebra.
}
\label{table1}
\end{table}

\section{General requirement for having a $n$th-order topological insulator in A and AIII classes}
According to the symmetry classification, A and AIII classes are the two complex classes without anti-unitary symmetries, distinguished by the absence and presence of a chiral symmetry \cite{schnyder2008classes,ryu2010class,RevModPhys.88.035005}. These two classes support odd and even nontrivial topological defects respectively, characterized by a $Z$ invariant \cite{teo2010classes}.
Following the main text, we consider a general $d$D Hamiltonian with $J$ anticommuting terms,
\begin{eqnarray}
H_d=\mathbf{h}(\mathbf{k})\cdot\mathbf{\gamma}=\sum_{j=1}^J h_j(\mathbf{k})\gamma_j,
\end{eqnarray}
with $\gamma_j$ satisfying $\{\gamma_j,\gamma_{j'}\}=2\delta_{jj'}$ and $\mathbf{k}=(k_1,k_2,...k_d,...,k_d)$. 
%The simplest example of $\gamma_j$ matrices is the three Pauli matrices $\sigma_{x,y,z}$. For $J>3$, $\gamma_j$ can be constructed by multiplying an extra set of Pauli matrices. That is, for $J$ anticommuting $\gamma_j$, $J+2$ anticommuting matrices can be obtained as $(\gamma_j\sigma_1, \gamma_0\sigma_2,\gamma_0\sigma_3)$, with $\sigma_{1,2,3}$ being the three Pauli matrices in an arbitrary order, and $\gamma_0$ an identity matrix.
The eigenenergies of this system are given by
\begin{eqnarray}
E_{\pm}=\sqrt{\sum_{j=1}^J h_j^2},
\end{eqnarray}
and the gap between $E_\pm$ closes when each $h_j(\mathbf{k})=0$. Therefore the system generally has some gapless points in the BZ when $J\leqslant d$, and a gapped phase can only be obtained with polarized pseudospin throughout the BZ (i.e. at least one $h_j$ is always positive or always negative). 
However, this gapped phase must be topologically trivial, as the gap is guaranteed by one nonzero $h_j$, so that all other terms can be smoothly tuned to zero without closing the gap.
Therefore the system is topologically equivalent to that with only the nonzero $h_j$ term, which is a topologically trivial band insulator, as the Hamiltonian ($h_j\gamma_j$) cannot give any nontrivial winding. 

For $J=d+1$, $\mathbf{h}(\mathbf{k})$ forms a $d$D closed manifold across the BZ, which encloses the origin of the $J$-dimension vector space an integer number of times, resulting in a quantized topological invariant. Consequently, 1st-order topological boundary states emerge under OBCs due to the topological bulk-boundary correspondence. 
By constructing the $\gamma_j$ matrices as products of different sets of Pauli matrices, we can see that the
absence and presence of a chiral symmetry are given by the parity of $J$, 
as the allowed number of anticommuting terms is always increased by $2$ when a new set of Pauli matrices is introduced.
One of the simplest examples is the Su-Schrieffer-Heeger (SSH) model \cite{SSH}, which is a 1D model described by two Pauli matrices ($J=2$), satisfying a chiral symmetry whose symmetry operator is given by the third Pauli matrix. In this model, $h_1(k)$ and $h_2(k)$ give a 1D loop in the 2D vector space, and the topological invariant is the winding number of the 1D loop regarding the origin of the vector space. Note that this model also satisfies time-reversal and particle-symmetries, and hence belongs to the BDI class. Nevertheless, it shares the same $Z$-type topological properties in 1D, and are both described by the winding number.

When $J$ is further increased, the vector space expands with extra dimensions, and the d$D$ manifold can no longer enclose its origin. In this scenario, the 1st-order boundary states are not topologically protected, and may merge into the bulk spectrum without a gap closing. 
On the other hand, upon opening up the boundary along a given spatial direction $x$, the existence of these 1st-order boundary states can be determined by a winding profile of the Hamiltonian vector with $k_x$ varying a period, which effectively corresponds to two dimensions of the vector space to define the winding. More explicitly, we may rotate and stretch the vector space, i.e. recombine and rescale different anticommuting terms, to have $k_x$ appear in only two terms, without changing the above mentioned winding topology. Through this process, the behavior of these 1st-order boundary states can be described by an effective $(d-1)$D Hamiltonian $H_{d-1}$ constructed by the rest $J-2$ anticommuting terms \cite{Mong2011winding,li2017engineering}. Suppose $H_{d-1}$ possesses a $n$th-order topology, it shall be inherited by the $(d-1)$D 1st-order boundary states of the original system, resulting in $(d-1-n)$D boundary states associated with the topology of $H_{d-1}$, which, by definition, are the $(n+1)$th-order boundary states of the original system. 
We can see that the existence of these $(n+1)$th-order boundary states depends on both the existence of the 1st-order boundary states of the original system, determined by a topological winding profile (the winding number $\nu_1$ in the main text); and the $n$th-order topology of $H_{d-1}$. Hence they are topologically protected, reflecting a $(n+1)$th-order topology of the overall system.

In the above discussion, if we let $n=1$, we can see that $H_{d-1}$ needs to have $d-1+1=d$ anticommuting terms to exhibit its intrinsic 1st-order topology. Thus $H_{d}$ needs to have $d+2$ anticommuting terms to exhibit a $2$nd-order topology, and so forth, $J=d+n$ anticommuting terms are required for the system to exhibit a ($Z$-type) $n$th-order topology.

\section{Quantum quench dynamics to detect pseudospin textures}
\subsection{Dynamical characterization of the 2rd-order topological insulator}
As mentioned in the main text, the winding behavior of pseudospin textures associated with a BIS at any nesting level can be observed by the time-averaged pseudospin texture in quantum quench dynamics \cite{zhang2018dynamical}.
Nevertheless, it generally requires more than one quenching process to obtain the full topological information of a given system \cite{zhang2018dynamical,yu2020high}.
Specifically, let us use the 3D 2nd-order topological insulator case as an example.
First, we need to locate the BISs of the system, which are given by zeros of different pseudospin components.
To this end, we consider an initial state $\psi_{\rm ini}$ as an eigenstate of the Hamiltonian $H_{\rm ini}=H(\mathbf{k})+m_{aa'}\sigma_a\tau_{a'}$ with $m_{aa'}$ much larger than the other parameters. This extra term polarizes the pseudospin along $\sigma_a\tau_{a'}$ direction.
The post-quench Hamiltonian is given by $H(\mathbf{k})$, and the time-averaged pseudospin texture of $\sigma_b\tau_{b'}$ after the quench is given by
\begin{eqnarray}
%\overline{\langle\sigma_a\tau_{a'}\rangle}=\frac{1}{T}\int_0^T\langle\psi_{\rm fin}|\sigma_a\tau_{a'}|\psi_{\rm fin}\rangle dt,
\overline{\langle\sigma_b\tau_{b'}\rangle}_{aa'}=\frac{1}{T}\int_0^T{\rm Tr}\left[\rho e^{iHt} \sigma_b\tau_{b'} e^{-iHt} \right] dt,\label{eq:dynamic}
\end{eqnarray}
with $\rho$ the density matrix of the initial state.
\GJB{Complementing discussions in the main text}, here we have further specified the direction of pre-quench pseudospin polarization ($aa'$) in the notation of $\overline{\langle\sigma_b\tau_{b'}\rangle}_{aa'}$.
The quenching dynamics leads to a procession of the pseudospin vector about the Hamiltonian vector $\mathbf{h}(\mathbf{k})$,
therefore when $\mathbf{h}(\mathbf{k})$ is perpendicular to the initial pseudospin polarization ($h_{aa'}(\mathbf{k})=0$),
a vanishing time-averaged pseudospin texture $\overline{\langle\sigma_b\tau_{b'}\rangle}_{aa'}=0$ shall be obtained over long-time dynamics.
Otherwise, nonzero time-averaged pseudospin texture emerges and points toward either the same or opposite direction of $\mathbf{h}(\mathbf{k})$, depending on the sign of $h_{aa'}(\mathbf{k})$.
Therefore the component $\langle\sigma_a\tau_{a'}\rangle_{aa'}$ of the time-averaged pseudospin texture, \blue{denoted as $\langle\sigma_a\tau_{a'}\rangle$ in the main text, }vanishes only at the momentum with $h_{aa'}(\mathbf{k})=0$,
and the BISs can be determined by the zeros of $\langle\sigma_a\tau_{a'}\rangle_{aa'}$. More explicitly,
$S_1^{m=2}$ and $S_2^{m=1}$ are given by the zeros of $(\langle\sigma_x\tau_{0}\rangle_{x0},\langle\sigma_z\tau_{0}\rangle_{z0})$ and $\langle\sigma_y\tau_{z}\rangle_{yz}$ respectively.
%Note that if $(a,a')\neq(b,b')$ in Eq. (\ref{eq:dynamic}), $\langle\sigma_b\tau_{b'}\rangle_{aa'}$ vanishes at the zeros of both $h_{aa'}$ and $h_{bb'}$,
On the other hand, $\langle\sigma_a\tau_{a'}\rangle_{aa'}$ can only take the same sign (or zero) due to the strong polarization of the initial state, therefore it does not reflect the sign of $h_{aa'}$ or the associated winding information.

In order to extract the winding information of the pseudospin texture associated with different BISs, one must consider different quenching processes. As discussed in the main text, the first winding number $\nu_1$ is defined for $(h_{x0},h_{z0})$ along a trajectory away from $S_1^{m=2}$. Therefore it can be detected by $(\langle\sigma_x\tau_{0}\rangle_{aa'},\langle\sigma_z\tau_{0}\rangle_{aa'})$ as long as $(a,a')\neq(x,0)$ or $(z,0)$, and the trajectory does not coincide with any zero of $h_{aa'}$.
The first condition is to have the time-averaged values to reflect the sign information of $h_{x0}$ and $h_{z0}$ (and hence the winding information).
The second condition is to ensure that $h_{aa'}$ does not change sign along the trajectory, because the sign of $h_{aa'}(\mathbf{k})$ determines whether the time-averaged pseudospin texture, obtained from the procession of the post-quench state, is parallel or anti-parallel to $\mathbf{h}(\mathbf{k})$.

Finally, the second winding number $C$ is associated with $(h_{yx},h_{yy})$ along $S_2^{m=1}$,
therefore we can obtain $C$ by looking at the winding of $(\langle\sigma_y\tau_{x}\rangle_{aa'},\langle\sigma_y\tau_{y}\rangle_{aa'})$ with similar restriction of $(a,a')$ as above.
In the main text, we have chosen $(a,a')=(x,0)$ as $h_{x0}$ is generally nonzero along $S_2^{m=1}$,
\blue{and denoted these time-averaged pseudospin textures \GJB{with the winding information} as $\langle\sigma_b\tau_{b'}\rangle'$, to distinguish from those that \GJB{locate} the BISs.}

\subsection{Dynamical characterization of the 3rd-order topological insulator}
Similar to the previous example of 2nd-order topological insulator, here we also need several steps to topologically characterize the system with dynamical properties.
First, the three BISs $S_1^{m=2}$, $S_2^{m=2}$, and $S_3^{m=1}$ are given by the zeros of the time-averaged pseudospin vector
\begin{eqnarray}
\bar{P}_1=(\overline{\langle\sigma_z\tau_{0} s_{0}\rangle}_{z00},\overline{\langle\sigma_x\tau_{0} s_{0}\rangle}_{x00}),\label{eq:P1}\\
\bar{P}_2=(\overline{\langle\sigma_y\tau_{z} s_{0}\rangle}_{yz0},\overline{\langle\sigma_y\tau_{x} s_{0}\rangle}_{yx0}),\label{eq:P2}
\end{eqnarray}
and $\overline{\langle\sigma_y\tau_{y} s_{x}\rangle}_{yyx}$ ($\overline{\langle\sigma_y\tau_{y} s_{x}\rangle}$ in the main text) respectively, with a different pre-quench polarizing direction for each pseudospin component.
Next, the two winding numbers $\nu_1$ and $\nu_2$ are determined by the winding of
\begin{eqnarray}
\bar{P}'_1=(\overline{\langle\sigma_z\tau_{0} s_{0}\rangle}_{yyx},\overline{\langle\sigma_x\tau_{0} s_{0}\rangle}_{yyx}),\label{eq:P11}\\
\bar{P}'_2=(\overline{\langle\sigma_y\tau_{z} s_{0}\rangle}_{yyx},\overline{\langle\sigma_y\tau_{x} s_{0}\rangle}_{yyx}),\label{eq:P22}
\end{eqnarray}
along the trajectories shown in Fig.~\ref{fig:dynamic_3rd} [$(k_x,k_y)=0$ in the 3D BZ for $\bar{P}'_1$, and $k_x=0$ in the 2D BZ for $\bar{P}'_2$].
The pre-quench polarizing direction is chosen as $yyx$, because $h_{yyx}\neq 0$ along these trajectories for the parameters we considered.

The third topological invariant $C$ is determined by a sign function related to the time-averaged values of the pseudospin components $\sigma_y\tau_{y} s_{x}$ and $\sigma_y\tau_{y} s_{z}$.
In order to obtain the required information, we consider a pre-quench Hamiltonian $H_{\rm ini}^{x00}=H(\mathbf{k})+m_{x00}\sigma_x\tau_0 s_0$,
so that the polarizing direction of initial state is not perpendicular to the Hamiltonian vector in the regime close to $S_3^{m=1}$.
With the same post-quench Hamiltonian $H(\mathbf{k})$, we obtain the quantities $\overline{\langle\sigma_y\tau_{y} s_{x}\rangle}_{x00}$ and $\overline{\langle\sigma_y\tau_{y} s_{z}\rangle}_{x00}$, denoted as $\overline{\langle\sigma_y\tau_{y} s_{x}\rangle}'$ and $\overline{\langle\sigma_y\tau_{y} s_{z}\rangle}'$ in the main text, and hence the topological invariant $C$.

\section{A 3D 2nd-order topological insulator with large $\nu_z$}
The system described by Eqs. (2) and (3) in the main text has two winding numbers $\nu_1$ up to $1$, and $C$ up to $3$. Here we consider a more complicated toy model where $\nu_1$ also takes a value larger than $1$. The system is described by the Hamiltonian $H(\mathbf{k})=H'_1(\mathbf{k})+H_2(\mathbf{k}_{\parallel})$
\begin{eqnarray}
H'_{1}&=&h'_{x0}\sigma_x\tau_0+h'_{z0}\sigma_z\tau_0\nonumber\\
H_2(\mathbf{k}_{\parallel})&=&(\lambda \sin 2k_y)\sigma_y\tau_x-\lambda\sin 2k_x\sigma_y\tau_y\nonumber\\
&&+\left[m_1-t_1(\cos k_x+\cos k_y)\right]\sigma_y\tau_z,\label{model2}
\end{eqnarray}
where $h'_{x0}=t_z^2\sin^2k_z-\left[t_0(\cos k_x+\cos k_y+\cos k_z)-m_0\right]^2$ and $h'_{z0}=t_z\sin k_z\left[t_0(\cos k_x+\cos k_y+\cos k_z)-m_0\right]$. The 2-BIS $S_1^{m=2}$ of this system is defined with $H'_1=0$, which takes the same shape as that of the model in the main text, but the regime it encloses has a winding number $\nu_1=-2$ \cite{li20172}. Consequently, the number of 1st-order boundary states related to $\nu_1$ is doubled, and so is that of 2nd-order topological hinge states.

\begin{figure}
\includegraphics[width=1\linewidth]{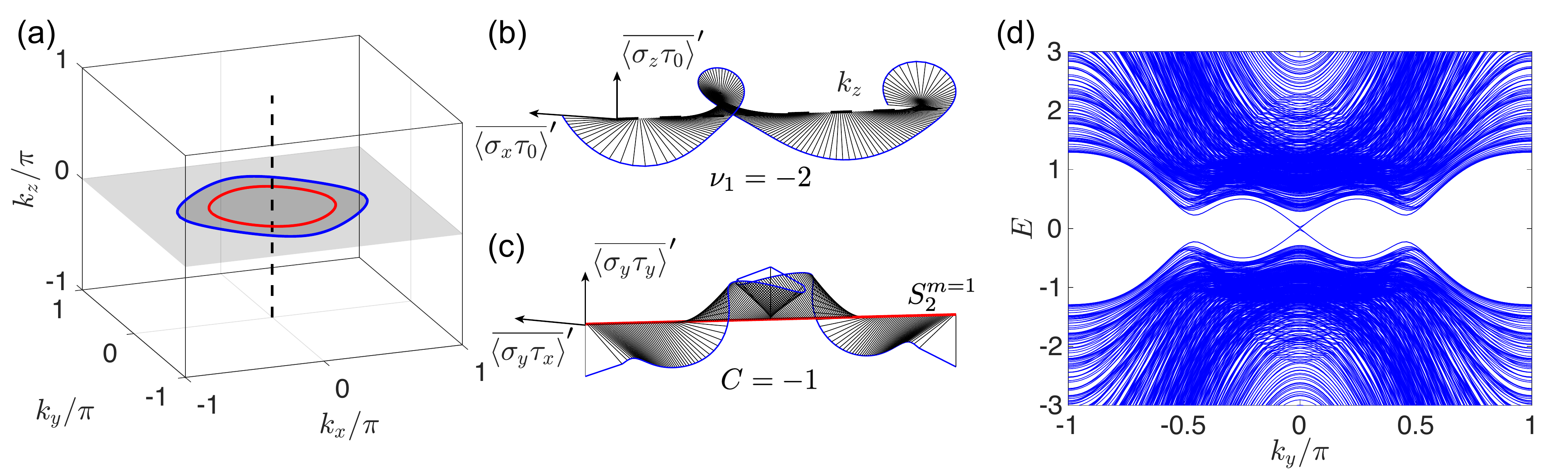}
\caption{
BISs and topological information of the 3D system with 2nd-order topology, described by the Hamiltonian Eq. (\ref{model2}). (a) the 3D Brilouin zone with a 2-BIS $S_1^{m=2}$ (blue loop) given by $H'_1=0$, and a nested 1-BIS $S_2^{m=1}$ (red loop) given by $h_{yz}=m_1-t_1(\cos k_x+\cos k_y)=0$ in the $k_z=0$ plane (gray) containing the $S_1^{m=2}$. The winding number $\nu_1$ is defined for $H'1$ with fixed $k_z$ varying a period at fixed $(k_x,k_y)$, e.g. along the dash line with $k_x=k_y=0$. Darker gray color indicates the regime with $\nu_1=-2$, and lighter one is with $\nu_z=0$. (b,c) the winding of time-averaged pseudospin texture for different pseudospin components, along the dash line and $S_2^{m=1}$ (red loop, with an infinitesimal shifting) in (a) respectively. 
The initial state is chosen to be polarized along $\sigma_y\tau_z$ direction for (b), and $\sigma_x\tau_0$ direction for (c). 
(d) $xz$-OBC spectrum of this system, the in-gap  connecting the two bands are 4-fold degenerate, two localized at the hinges of the top surface along $z$ direction, and the other two at the hinges of the bottom.}
\label{fig:high_nuz}
\end{figure} 

In Fig. \ref{fig:high_nuz}, we illustrate an example of this model with $C=1$ for the nested 1-BIS $S_2^{m=1}$ (red loop) lying within the topologically nontrivial regime of the 2-BIS $S_1^{m=2}$ (dark region enclosed by the blue loop). Following the main text, we quench an eigenstate of a pseudospin-polarized system of $H_{\rm ini}=H(\mathbf{k})+m_{xz}\sigma_y\tau_z$ to a topologically nontrivial one, and the winding of different pseudospin textures along different trajectories in Fig. \ref{fig:high_nuz}(a) reflects the topological invariant of $\nu_1$ and $C$ respectively, as shown in Fig. \ref{fig:high_nuz}(b,c). 
Note that the time-averaged psudospin textures associated with $C$ need to be measured along a trajectory slightly shifted away from $S_2^{m=1}$, as discussed in the main text.
For the parameters we choose, the $xz$-OBC spectrum of this model has four pairs of gapless hinge states, which are degenerate in energy. Among them, two pairs are localized at the top surface, and the other two at the bottom. Thus on each surface, the number of pairs of hinge states is given by $\nu_1\times C$,  consistent with the results in the main text.

\section{Crossed band inversion surfaces}
In the 3D 2nd-order topological insulating system we consider in the main text, $S_1^{m=2}$ of the overall system and $S_2^{m=1}$ of the subsystem are both centered at $k_x=k_y=0$, thus one of them always enclosing the other, if they do not overlap. In more general cases, the two BISs may also cross each other at several 0D points, and an alternative approach is required to unveil their topological properties. 

\begin{figure}
\includegraphics[width=0.8\linewidth]{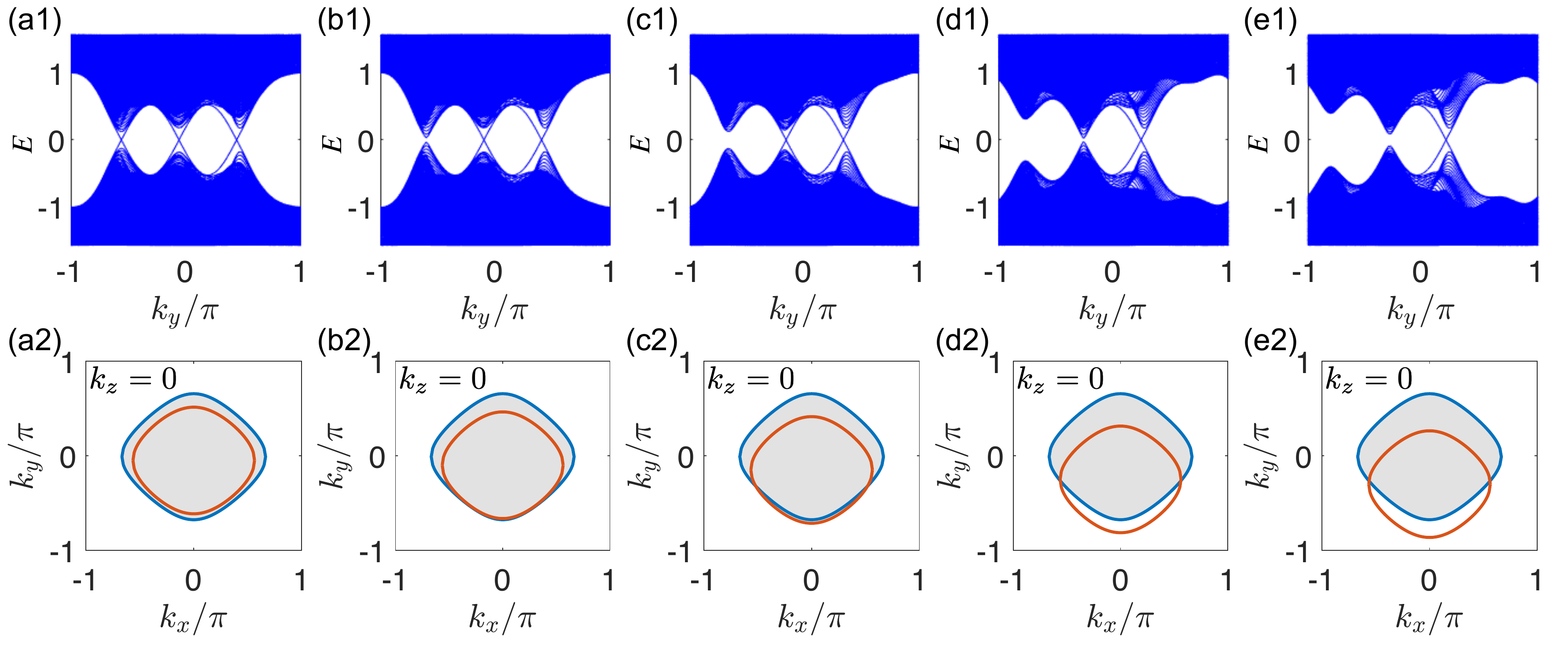}
\caption{(a1)-(e1) $xz$-OBC spectra and (a2)-(e2) BISs of the system described by Eqs. (\ref{eq:shift_1}) and (\ref{eq:shift_2}). Blue and red loops represent the $S_1^{m=2}$ and $S_2^{m=1}$, a 2-BIS of the whole system given by $H_1=0$ and a 1-BIS of $H'_2$ given by $h_{yz}=0$ respectively. 
Shadowed regimes correspond to a winding number $\nu_1=1$ associated with $S_1^{m=2}$.
The momentum shifting of $S_2^{m=1}$ is given by $\phi=0.05\pi$, $0.1\pi$, $0.15\pi$, $0.25\pi$, and $0.1\pi$ from left to right.  
Other parameters are $t_0=t_1=t_z=1$, $m_0=1.5$, $m_1=0.8$, and $\lambda=0.5$, which give $C=3$ for the subsystem $H'_2$.
%\red{[To be changed to cases with more clear touching BISs.]}
}
\label{fig:crossing}
\end{figure}

Consider the 3D 2nd-order topological system discussed in the main text with a shift of $k_y$ in $H_2$, described by the Hamiltonian $H(\mathbf{k})=H_1(\mathbf{k})+H_2(\mathbf{k}_{\parallel})$
with
\begin{eqnarray}
H_1(\mathbf{k})&=&\left[t_0(\cos k_x+\cos k_y+\cos k_z)-m_0\right]\sigma_x\tau_0\nonumber\\
&&+t_z\sin k_z\sigma_z\tau_0,\label{eq:shift_1}\\
H'_2(\mathbf{k}_{\parallel})&=&\lambda \sin 2k_y'\sigma_y\tau_x-\lambda\sin 2k_x\sigma_y\tau_y\nonumber\\
&&+\left[m_1-t_1(\cos k_x+\cos k_y')\right]\sigma_y\tau_z,\label{eq:shift_2}
\end{eqnarray}
and $k_y'=k_y+\phi$. The parameter $\phi$ only shifts $k_y$ for $H'_2$, which does not change the topological invariants defined for $H_1$ and $H'_2$ respectively.
By tuning $\phi$ away from zero, $S_2^{m=1}$ of the system will move along $k_y$ axis, and cross $S_1^{m=2}$ at some points. In these cases, we can see that the number of 2nd-order topological boundary states varies with $\phi$, and only partially reflects the topological invariant $C$ of $S_2^{m=1}$, as shown in Fig.~\ref{fig:crossing}. 
%\red{[Note: the figure is to be replaced by a new one]}

To determine the phase transition point, we note that topological properties of the effective 2D Hamiltonian $H'_2$ do not change with different $\phi$. For the parameters we choose, $H'_2$ itself always corresponds to three pairs of topological boundary states. However, for the overall 3D system, these topological properties are reflected by the 1st-order surface states upon $z$-OBCs, which present only within the nontrivial regime of $S_1^{m=2}$ with $\nu_1\neq 0$. 
Therefore, a pair of topological boundary states of $H_2'$ may manifest as gapless hinge states of the 3D system, only when its crossing point falls within the region of momentum where 1st-order surface states exist. 

In a 2D Chern insulating system of $H_2'$, the crossing points of its topological boundary states can also be determined through BIS analysis. To see this, we first define a 2-BIS of $H_2'$ as $\lambda\sin 2k_x=m_1-t_1(\cos k_x+\cos k_y')=0$. This 2-BIS is made of some 0D points in the 2D BZ. Upon OBCs along $x$ direction, the projections of the 2-BIS separate the 1D edge BZ into regimes with and without edge states, which can be characterized by a winding number $\nu_y$ defined for each $k_y$. The eigenenergies of these 1D edge states are directly given by the third term $\lambda \sin 2k_y'$, hence the they can be degenerate at zeros only when $\lambda \sin 2k_y'=0$. Similarly, crossing points of edge states under OBCs along $y$ direction fall at $\lambda\sin 2k_x=0$, which can be seen by defining another 2-BIS as $\lambda \sin 2k_y'=m_1-t_1(\cos k_x+\cos k_y')=0$.

\begin{figure}
\includegraphics[width=0.6\linewidth]{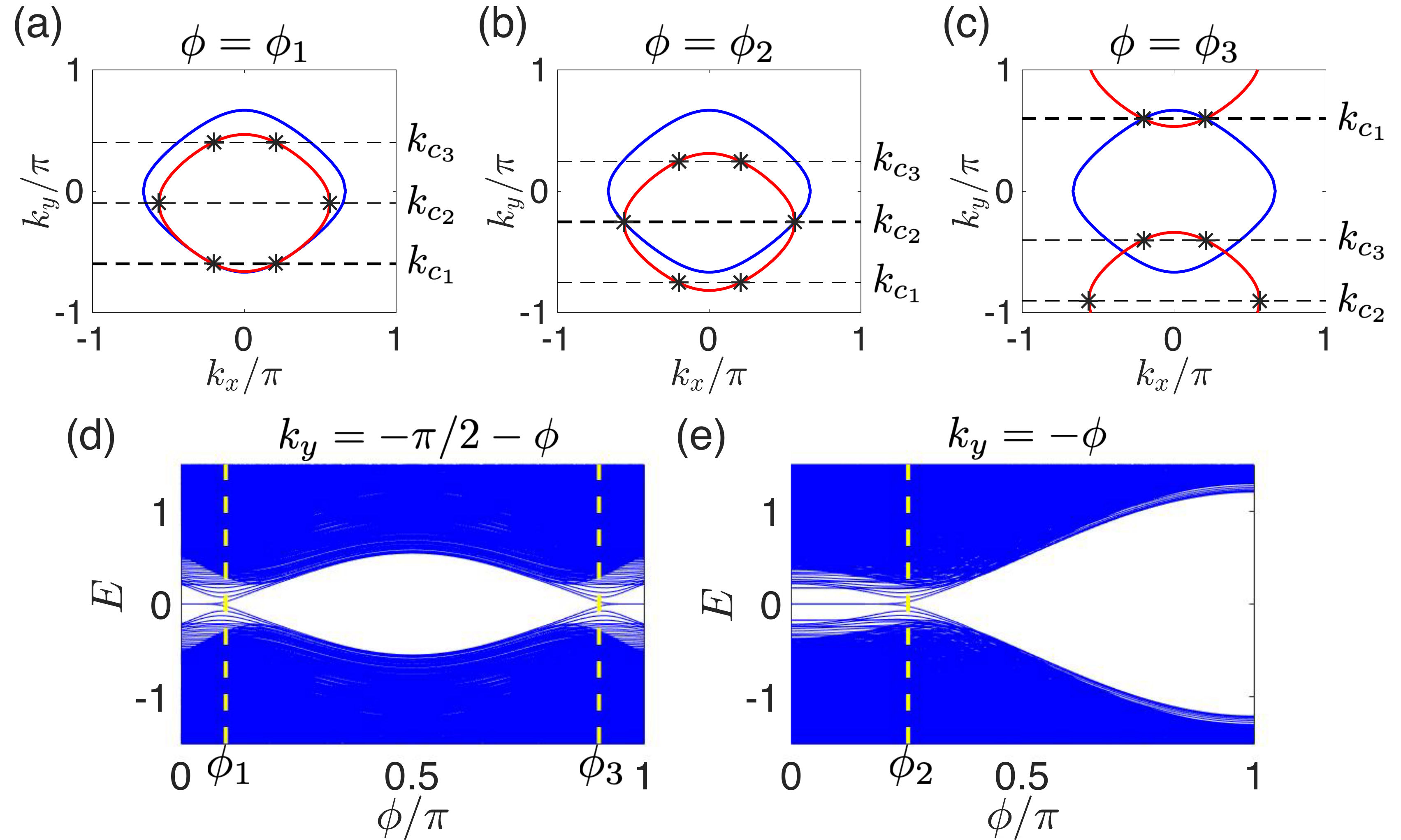}
\caption{
(a)-(c) BISs of the system described by Eqs. (\ref{eq:shift_1}) and (\ref{eq:shift_2}) with different momentum shifting $\phi$ of $H_2'$.
Blue and red loops indicate the BISs $S_1^{m=2}$ and $S_2^{m=1}$ respectively.
The black dash lines are given by three solutions of $\lambda \sin 2(k_y+\phi)=0$, $k_y\in(-\pi,\pi]$, with $k_{c_1}=-\pi/2-\phi$, $k_{c_2}=-\pi/2$, and $k_{c_3}=\pi/2-\phi$.
The crossing points between them and $S_2^{m=1}$ gives a 2-BIS $S_2^{m=2}$ (black stars) of the subsystem $H_2'$.
In each panel, $S_2^{m=2}$ coincides with $S_1^{m=2}$ at two of the six points, indicating a topological phase transition.
(d)-(e) the $xz$-OBC spectra at $k_y=k_{c_1}=-\pi/2-\phi$ and $k_y=k_{c_2}=-\phi$ respectively, with $\phi$ varying from $0$ to $\pi$.
Topological phase transitions are seen to occur at $\phi=\phi_{1,2,3}$, where different stars of $S_2^{m=2}$ coincide with $S_1^{m=2}$ as shown in (a)-(c).
The system's size is chosen as $N_x=40$, $N_z=60$ for (d) and (e).
Other parameters are $t_0=t_1=t_z=1$, $m_0=1.5$, $m_1=0.8$, and $\lambda=0.5$.
}
\label{fig:transition}
\end{figure} 

In the above 3D model, we have kept PBCs along $y$ direction, thus $k_y$ is always a good quantum number, and the crossing points of 2nd-order boundary states satisfy 
\begin{eqnarray}
\lambda \sin 2k_y'=\lambda \sin 2(k_y+\phi)=0. 
\end{eqnarray}
Together with the 1-BIS $S_2^{m=1}$, this condition gives a 2-BIS $S_2^{m=2}$ of the subsystem $H_2'$, as shown by the three pairs of black stars in Fig. \ref{fig:transition}(a)-(c).
Thus the (1st-order) topological properties of $H_2'$ can also be captured by the pseudospin texture on $S_2^{m=2}$,
and we may expect a topological phase transition to occur when each pair of the stars of $S_2^{m=2}$ moves outside $S_1^{m=2}$. 
For the parameters we choose, these transition points are given by
\begin{eqnarray}
\phi_1&=&\arccos{(m_0/t_0-1-m_1/t_1)}-\pi/2\approx0.097\pi,\nonumber\\
\phi_2&=&\arccos{(m_0/t_0-m_1/t_1)}\approx0.253\pi,\nonumber\\
\phi_3&=&3\pi/2-\arccos{(m_0/t_0-1-m_1/t_1)}\approx0.903\pi,\nonumber\\
\end{eqnarray}
when $\phi\in [0,\pi]$, as shown in Fig. \ref{fig:transition}. We can see that the 2nd-order topological properties of the overall system are determined by the parts of $S_2^{m=2}$ falling within the nontrivial regime of $S_1^{m=2}$ ($\nu_1=1$), consistent with the method of nested BISs we proposed in the main text.

%\clearpage
%

\end{document}